\documentclass[prb,superscriptaddress,twocolumn,floatfix]{revtex4}
\usepackage{graphicx}
\usepackage{color}
\usepackage{dcolumn}
\usepackage{amsmath}
\usepackage{amssymb}

\newcommand{\fstx}{FeSe$_x$Te$_{1-x}$}
\newcommand{\fst}{FeSe$_{0.5}$Te$_{0.5}$}

\begin{document}

\title{Coupling of spin and orbital excitations in the iron-based superconductor FeSe$_{0.5}$Te$_{0.5}$}
\author{S.-H. Lee}
\affiliation{Department of Physics, University of Virginia, Charlottesville, VA 22904-4714}
\author{Guangyong Xu}
\author{W. Ku}
\affiliation{Condensed Matter Physics \&\ Materials Science Department, Brookhaven National Laboratory, Upton, New York 11973-5000}
\author{J. S. Wen}
\affiliation{Condensed Matter Physics \&\ Materials Science Department, Brookhaven National Laboratory, Upton, New York 11973-5000}
\affiliation{Department of Materials Science \&\ Engineering, Stony Brook University, Stony Brook, NY 11794}
\author{C. C. Lee}
\affiliation{Condensed Matter Physics \&\ Materials Science Department, Brookhaven National Laboratory, Upton, New York 11973-5000}
\author{N. Katayama}
\affiliation{Department of Physics, University of Virginia, Charlottesville, VA 22904-4714}
\author{Z. J. Xu}
\affiliation{Condensed Matter Physics \&\ Materials Science Department, Brookhaven National Laboratory, Upton, New York 11973-5000}
\affiliation{Physics Department, The City College of New York, New York, NY 10031}
\author{S. Ji}
\affiliation{Department of Physics, University of Virginia, Charlottesville, VA 22904-4714}
\author{Z. W. Lin}
\author{G. D. Gu}
\author{H.-B. Yang}
\author{P. D. Johnson}
\author{Z.-H. Pan}
\author{T. Valla}
\affiliation{Condensed Matter Physics \&\ Materials Science Department, Brookhaven National Laboratory, Upton, New York 11973-5000}
\author{M. Fujita}
\affiliation{Institute for Materials Research, Tohoku University, Senda, Miyagi 980-8577, Japan}
\author{T. J. Sato}
\affiliation{Neutron Science Laboratory, Institute for Solid State Physics, University of Tokyo, 106-1 Shirakata, Tokai, Ibaraki 319-1106, Japan}
\author{S. Chang}
\affiliation{NIST Center for Neutron Research, National Institute of Standards and Technology, Gaithersburg, MD 20899}
\author{K. Yamada}
\affiliation{WPI Research Center, Advanced Institute for Materials Research, Tohoku University, Sendai~980-8577, Japan}
\author{J. M. Tranquada}
\affiliation{Condensed Matter Physics \&\ Materials Science Department, Brookhaven National Laboratory, Upton, New York 11973-5000}

\date{\today}

\begin{abstract}
We present a combined analysis of neutron scattering and photoemission measurements on superconducting \fst.  The low-energy magnetic excitations disperse only in the direction transverse to the characteristic wave vector $(\frac12,0,0)$, whereas the electronic Fermi surface near $(\frac12,0,0)$ appears to consist of four incommensurate pockets.  While the spin resonance occurs at an incommensurate wave vector compatible with nesting, neither spin-wave nor Fermi-surface-nesting models can describe the magnetic dispersion.  We propose that a coupling of  spin and orbital correlations is key to explaining this behavior.  If correct, it follows that these nematic fluctuations are involved in the resonance and could be relevant to the pairing mechanism.
\end{abstract}

\maketitle

The quest to understand the mechanism of high temperature superconductivity gained new momentum with the recent discovery of Fe-based superconductors.\cite{hoso09}  Experimental studies of these materials have shown that by chemically tuning the carrier density one can obtain both magnetically ordered and superconducting phases.\cite{ishi09}  The superconducting transition temperature is maximized for conditions close to where magnetic order is suppressed, leading to predictions that magnetic fluctuations are important for electron pairing and motivating comparisons with copper-oxide superconductors.  Initial theories have treated the magnetic excitations as independent of the atomic orbital character of the conduction electron states,\cite{mazi08,kuro08} an assumption that seems to work well in other metallic magnetic systems; however, there have been recent proposals that coupled spin and orbital order occur in the antiferromagnetic state.\cite{krug09,lee09,lv09,turn09,cric09,chen09}   

In this paper, we present a study of low-energy magnetic and electronic excitations in superconducting \fst.  Particularly striking is the anomalous anisotropic dispersion of magnetic excitations, which is quite distinct from the spin-wave-like excitations typically seen in magnetic metals.  Angle-resolved photoemission spectroscopy (ARPES) reveals an unexpected splitting of larger Fermi-surface pockets into smaller pockets with distinct orbital characters.  A simple nesting picture does not appear to be compatible with the magnetic dispersion.  We propose, instead, that a strong coupling of spin and orbital correlations can lead to the observed anisotropy.  Such a coupling would suggest that electronic nematic fluctuations may be present in the normal state and may participate in the superconductivity.  This would be a rather different way to connect with the cuprates.\cite{frad09}  

\begin{figure}[t]
\center{\includegraphics[width=0.95\columnwidth,angle=0]{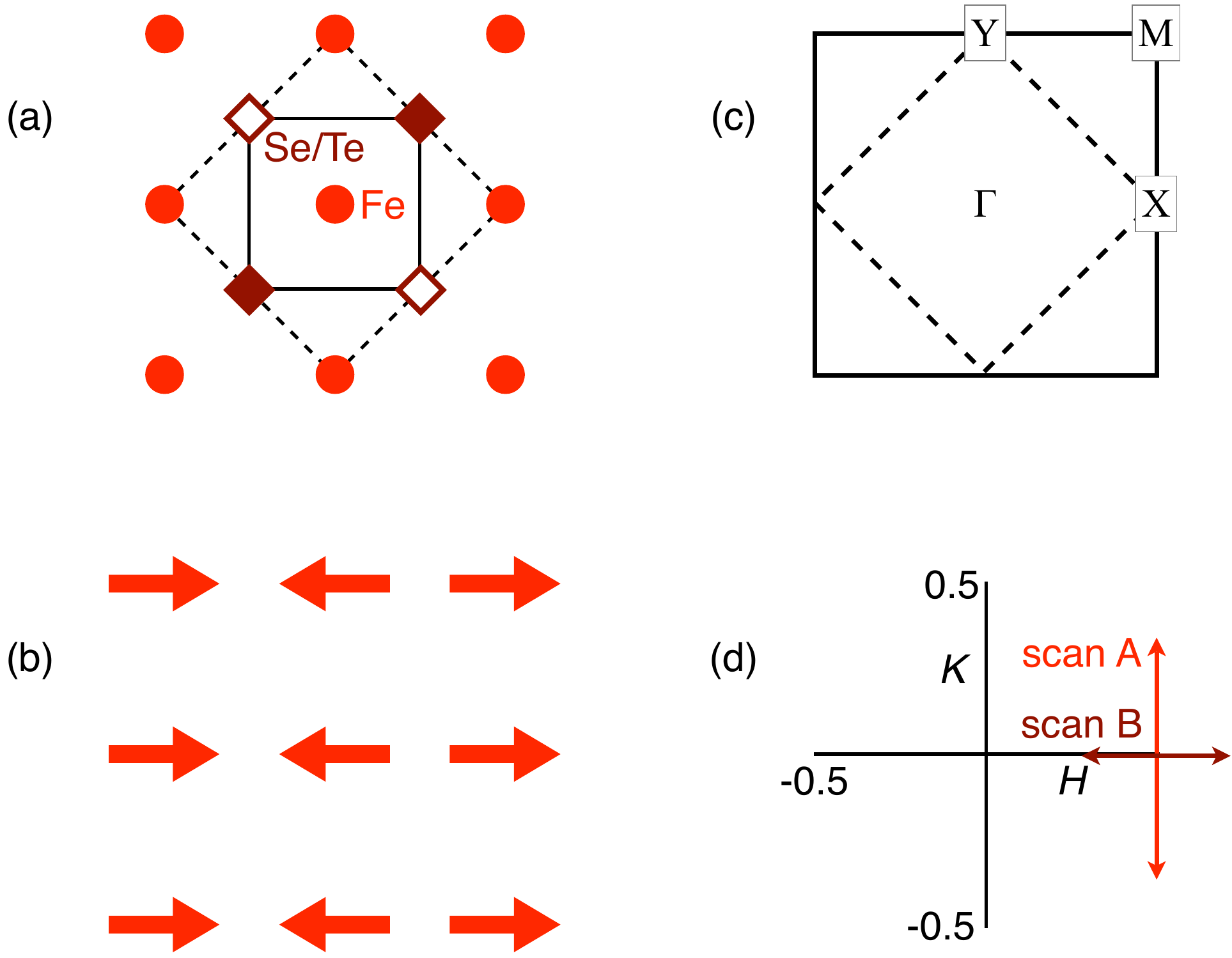}}
\caption{(color online). (a) Atomic positions within an Fe plane of \fst.  Filled (open) diamonds indicate Se/Te atoms that sit above (below) the plane.  Solid line: unit cell used here; dashed line: crystallographic unit cell. (b) Spin pattern corresponding to ${\bf Q} = (\frac12,0)$.  (c) Reciprocal space corresponding to (a), with symmetry point labeled.  Solid line: Brillouin zone for one Fe per unit cell; dashed line: Brillouin zone for two Fe per unit cell.  (d) Reciprocal space labeled in reciprocal lattice units, indicating directions of scans for the neutron scattering measurements.} \label{fg:latt}
\end{figure}

Before continuing, let us first resolve our choice of coordinates.  The structure of \fst\ contains layers of Fe atoms forming a square lattice, with Se/Te atoms centered above or below these squares, alternating in a checkerboard fashion [Fig~\ref{fg:latt}(a)].  The c-axis displacements of the Se/Te atoms break the translational symmetry, doubling the unit cell size; however, as we will show, this effect is rather weak for the low-energy electronic structure involving Fe-derived orbitals.  Hence, we choose to work with a unit cell containing one Fe atom ($a=b=2.69$~\AA\ and $c = 6.27$~\AA).  Reciprocal lattice vectors will be specified in units of $(2\pi/a,2\pi/a, 2\pi/c)$.

It is already experimentally established that the low energy magnetic excitations in superconducting \fstx\ occur near the in-plane wave vector ${\bf Q}=(\frac12,0)$,\cite{qiu09} which is the same as the magnetic ordering wave vector found in $R$FeAsO ($R=$ La, Ce, Pr, Nd) and $A$Fe$_2$As$_2$ ($A=$ Ca, Sr, Ba),\cite{lynn09} where it corresponds to columnar antiferromagnetic order [Fig.~\ref{fg:latt}(b)]. (Note that the magnetic ordering in Fe$_{1+\delta}$Te is different.\cite{bao09,li09})    Several recent theoretical analyses have proposed that orbital ordering, involving the  Fe $3d_{xz}$ and $3d_{yz}$ states, is an essential feature of the magnetically-ordered state.\cite{krug09,lee09,lv09,turn09,cric09,chen09}  This is consistent with the fact that a symmetry-lowering structural transition, which breaks the degeneracy of the $d_{xz}$ and $d_{yz}$ states, always occurs before magnetic ordering.\cite{lynn09}  Very recent experiments emphasize the electronic anisotropy of this state,\cite{fern09,nand10,chu09} demonstrating its nematic character \cite{fang08,chua10}.

For the experimental studies, single crystals of \fst\ were grown at Brookhaven by a unidirectional solidification method.  The nominal composition had no excess Fe, and $T_c$, measured by magnetic susceptibility, is 14~K.\cite{wen09}  The ARPES spectra were recorded on beamline U13UB at the National Synchrotron Light Sounce using a Scienta SES2002 electron spectrometer.  The incident photon energy was 17 eV, while the photoelectron energy and angular resolution was 15~meV and 0.1$^\circ$, respectively.  Samples (small pieces from the crystal studied with neutrons) were cleaved {\it in situ} (base pressure of $5\times 10^{-11}$ Torr) and held at $T=15$~K for the measurements.    Complementary ARPES data were collected on beamline 12.0.1 at the Advanced Light Source (ALS), using 50-eV photons (energy resolution of 25 meV) and $T=15$ K.

The neutron scattering experiments were performed at the cold-neutron triple-axis spectrometer SPINS located at the National Institute of Standards and Technology Center for Neutron Research (NCNR). The 9-g crystal was aligned in the $(hk0)$ scattering plane; a mosaic width of $<0.6^\circ$ (limited by resolution) was measured. The incident neutron energy was selected with a vertically-focusing pyrolytic graphite (PG) (002) monochromator. Scattered neutrons were analyzed with five 2.1 cm $\times$ 15 cm PG(002) flat analyzer blades that reflected neutrons with $E_f = 5.0$ meV onto a $^3$He proportional counter. A cooled Be filter was placed between the sample and analyzer to suppress higher order neutron contamination. 

\begin{figure}[t]
\center{\includegraphics[width=1\columnwidth,angle=0]{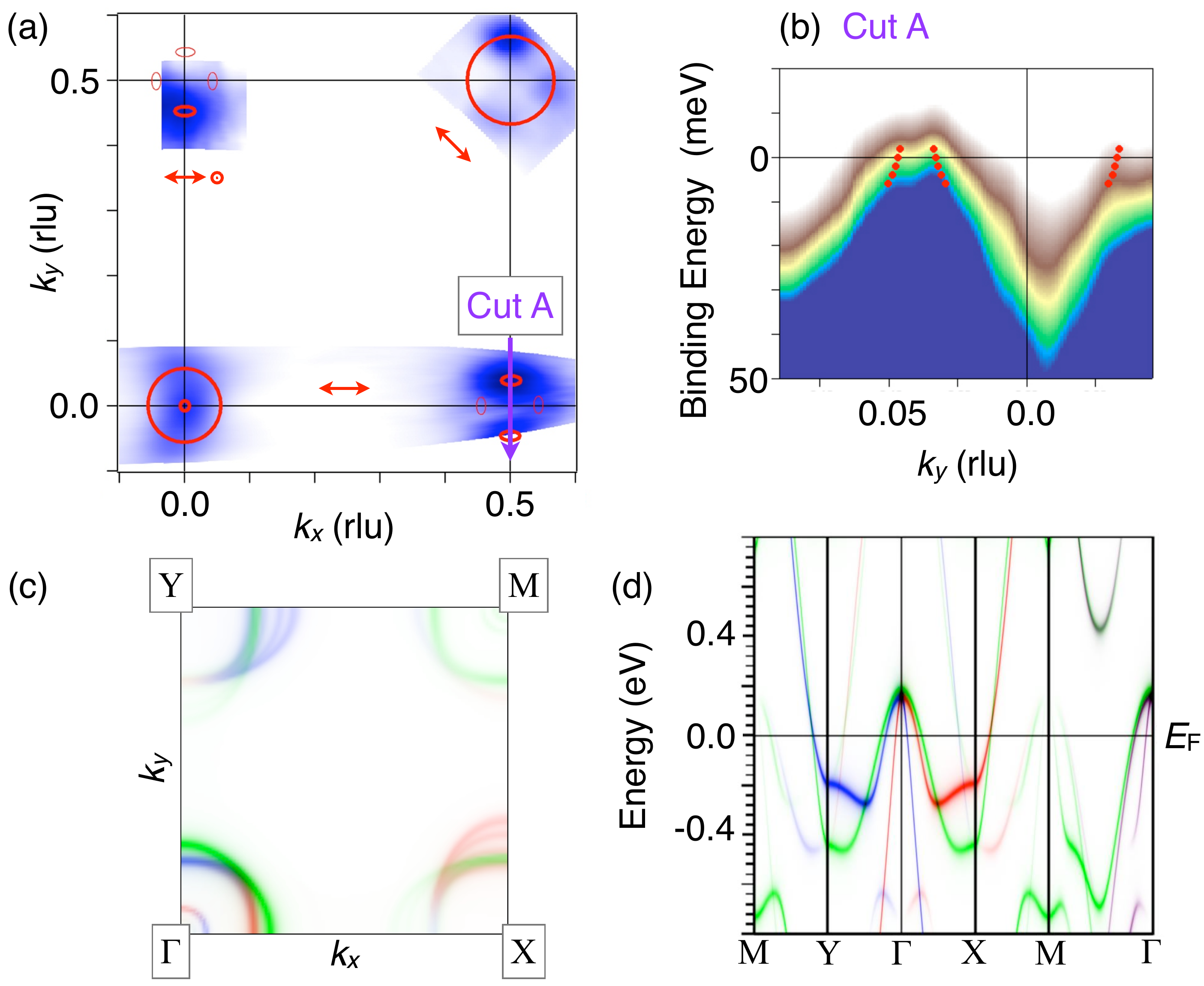}}
\caption{(a) ARPES intensity map at $E_F$, integrated over 0--10 meV binding energy; upper half measured at the ALS with 50-eV photons, lower half from the NSLS with 17-eV photons.  Thick red lines indicate estimated Fermi-surface positions; thin red lines denote pockets inferred by symmetry.  Red arrows denote polarization direction;  at $(0,0.5)$ polarization has a $k_z$ component that varies with $k_x$.  (b)  Intensity map (white = minimum, blue = maximum) along line denoted as Cut A in (a).  Red dots indicate positions of bands as they cross $E_F$.  Asymmetry about $k_y=0$ is due to imperfect sample orientation.  (c)  Calculated spectral weight at $E_F$, averaged over $k_z$.   Orbital coding:  red--$d_{yz}$; blue--$d_{xz}$; green--all others.  (d) Calculated band dispersions along symmetry directions, with intensity scaled according to spectral weight, obtained by unfolding bands calculated in a reduced zone\cite{ku10}; color code same as (c).} \label{fg:ph}
\end{figure}

Let us consider the ARPES results first. The electronic spectral weight within 0--10 meV of the Fermi energy, $E_F$, is plotted in Fig.~\ref{fg:ph}(a); the positions of identified band crossings, determined from energy-momentum cuts as in Fig.~\ref{fg:ph}(b), are indicated by thick red lines.  For comparison, our local density calculation of the Fermi surface is shown in Fig. 2(c) and the back-folded band structure in Fig. 2(d).  Several features are obvious.  First of all, we can see that the effect of the unit cell doubling due to the Se/Te positions is small.  To appreciate this, note from Fig.~\ref{fg:latt}(c) that for the doubled unit cell the $\Gamma$ and M points correspond to the same crystal momentum.  While the band dispersions must be the same at both points, the spectral weights of the individual bands need not be,\cite{ku10} and that is certainly the case here.  Indeed, in Fig.~\ref{fg:ph}(d) one can barely recognize the relationship between M and $\Gamma$ points, reflecting weak Umklapp effects on the Fe-orbitals.

Secondly, experiment indicates the presence of small, incommensurate pockets about the X and Y points, in contrast to the larger commensurate pockets predicted by our calculation.  The spectral intensities of these pockets are quite sensitive to the orientation of the photon polarization.  We can understand this dependence as follows.  The theoretical pockets about X (and Y) have distinct orbital character, as indicated in Fig.~\ref{fg:ph}(c) and (d).  Fe $d_{yz}$ character (red) dominates near $(\frac12,\pm k_0,0)$, $d_{xz}$ (blue) near $(\pm h_0,\frac12,0)$, and $d_{xy}$ around $(\frac12-h_0,0,0)$ and $(0,\frac12-k,0)$.  Due to the broken cubic symmetry at the Fe sites and the hybridization with the neighboring Te/Se $p$ orbitals, $d_{yz}$/$d_{xz}$/$d_{xy}$ Wannier orbitals actually contain $p_x$/$p_y$/$p_z$ character (see Ref.~\onlinecite{lee09} for a related analysis of Wannier functions).  Assuming that the ARPES matrix elements are primarily $p\rightarrow s$ (which is reasonable for low photon energies), the association between orbital character and polarization direction is: $d_{yz}$ : [100], $d_{xz}$ : [010], $d_{xy}$ : [001].  From the observed intensities, we conclude that the pockets at $(\frac12,\pm k_0,0)$ are $d_{yz}$-like, those at $(\pm h_0,\frac12,0)$ are $d_{xz}$-like, and those at $(0,\frac12\pm k_0,0)$ [and, by symmetry, at $(\frac12\pm h_0,0,0)$] are $d_{xy}$-like.  The orbital character of the spectral weight at $E_F$ is qualitatively compatible with calculations.\cite{ragh08,lee09}  At the same time, the observed small hole pockets are different from the larger Fermi surfaces of the local density calculation, probably due to the inability of this mean-field technique to properly account for dynamical magnetic correlations.

\begin{figure}[t]
\center{\includegraphics[width=0.999\columnwidth,angle=0]{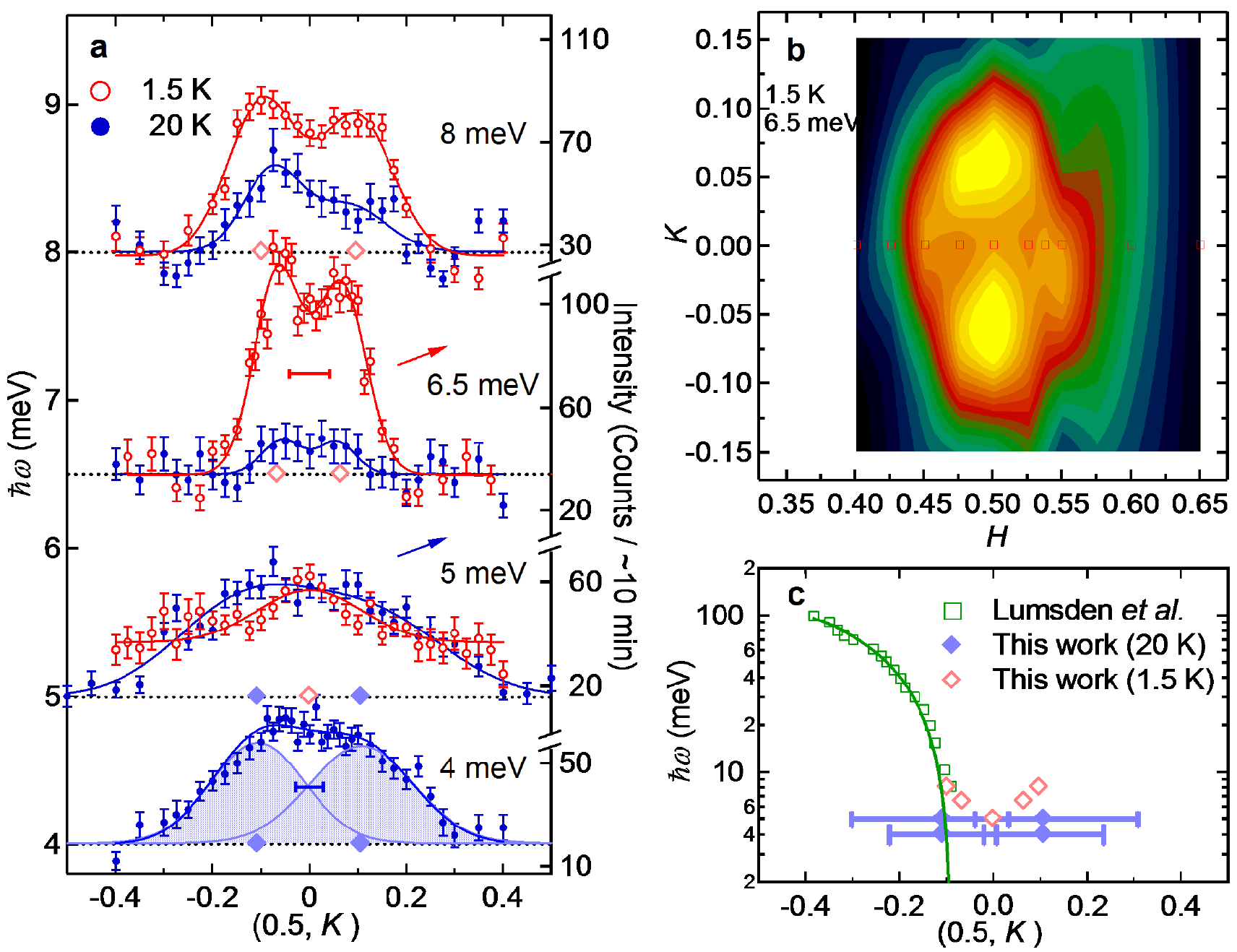}}
\caption{(color online). 
Momentum dependence of the spin resonance and low energy spin fluctuations. (a) Inelastic neutron scattering intensity as a function of {\bf Q}, obtained with constant energies, 8 meV, 6.5 meV, 5 meV, and 4 meV, at 1.5 K $(< T_c)$ (open red circles) and 20 K $(> T_c)$ (filled blue circles), taken along the transverse $(0.5,k)$ direction. The lines are fits to Gaussians; diamonds (open, 1.5~K; filled, 20~K) indicate fitted peak positions. Horizontal bars represent the full-width-of-the-half-maximum of the instrumental $Q$-resolution.  (b) Color contour map summarizing 1.5-K data obtained from a series of $(0.5,k)$ scans centered at positions marked by red squares. (c) Dispersion of the spin fluctuations. Open green squares extending to high energies are taken from Ref.~\onlinecite{lums10}, while open red  diamonds (1.5~K) and filled blue diamonds (20~K0 at low energies are our data taken from (a); horizontal bars are fitted peak widths at the lowest energies and $T=20$~K.  The line is explained in the text.
} \label{fg:ns2}
\end{figure}

Now let us turn to the neutron scattering results.  As already mentioned, a spin resonance has been observed to develop below $T_c$ in \fstx\ with $x=0.4$ at the in-plane wave vector ${\bf Q}=(\frac12,0)$ and an energy of 6.5~meV.\cite{qiu09}  We have confirmed the presence of a resonance in our $x=0.5$ sample.  What we wish to focus on here is the {\bf Q} dependence of the magnetic response in the vicinity of the resonance.  In the superconducting phase ($T=1.5$~K $<<T_c$), transverse scans along $(0.5,k)$ [scan A in Fig.~\ref{fg:latt}(d)] exhibit pairs of peaks at finite $|k|$ that disperse with energy above 5 meV, as shown in Fig.~\ref{fg:ns2}(a). The color-coded map of intensity vs.\ {\bf Q} at the resonance energy of 6.5 meV, Fig.~\ref{fg:ns2}(b), demonstrates an intriguing anisotropy: the transverse peaks are not reproduced along the longitudinal $(0.5+h,0)$ direction as one would expect from Fermi surface nesting between a pocket at $\Gamma$ and the four pockets about X. 

In the normal state ($T=20$~K), the spectral weight of the resonance is moved to lower energies, and the peaks appear to remain split down to 4~meV [blue symbols in Fig.~\ref{fg:ns2}(a)]; the asymmetry between transverse and longitudinal directions remains the same as below $T_c$. The dispersion of the peaks along the transverse direction is plotted in Fig.~\ref{fg:ns2}(c), with points at 10~meV and above from Lumsden {\it et al.}\cite{lums10}  (For similar recent results, see Refs.~\onlinecite{argy09,mook09b}.) The line through the points is a fit to $\hbar\omega = E_0 \sin (\pi (k- k_0))$ with $E_0 = 121$ meV for $k<k_0 = -0.09$. If these excitations were like spin waves, as in CaFe$_2$As$_2$,\cite{dial09,zhao09} we would expect to see cone-shaped dispersions coming out of $(0.5,\pm k_0)$, which is clearly not the case.  The dispersion of isolated intensity peaks along a single direction is quite unusual and requires consideration of factors beyond the degree of electronic correlation.

We propose that the solution to this puzzle involves the coupling of spin and orbital correlations.  As previously mentioned, such a coupling has already been proposed for the antiferromagnetic phase.\cite{krug09,lee09,lv09,turn09,cric09,chen09}  In fact, it has been shown\cite{lee09} that the reduction of lattice symmetry can be associated with long-range orbital ordering, and the resulting anisotropy of the superexchange parameters is consistent with the observed spin-wave dispersions.\cite{zhao09,dial09}  In the present case, the orbital degeneracy of $d_{xz}$ and $d_{yz}$ orbitals re-activates the orbital freedom in the absence of long-range orbital order.  This changes entirely the nature of the low-energy spin excitations.

\begin{figure}[t]
\center{\includegraphics[width=0.7\columnwidth,angle=0]{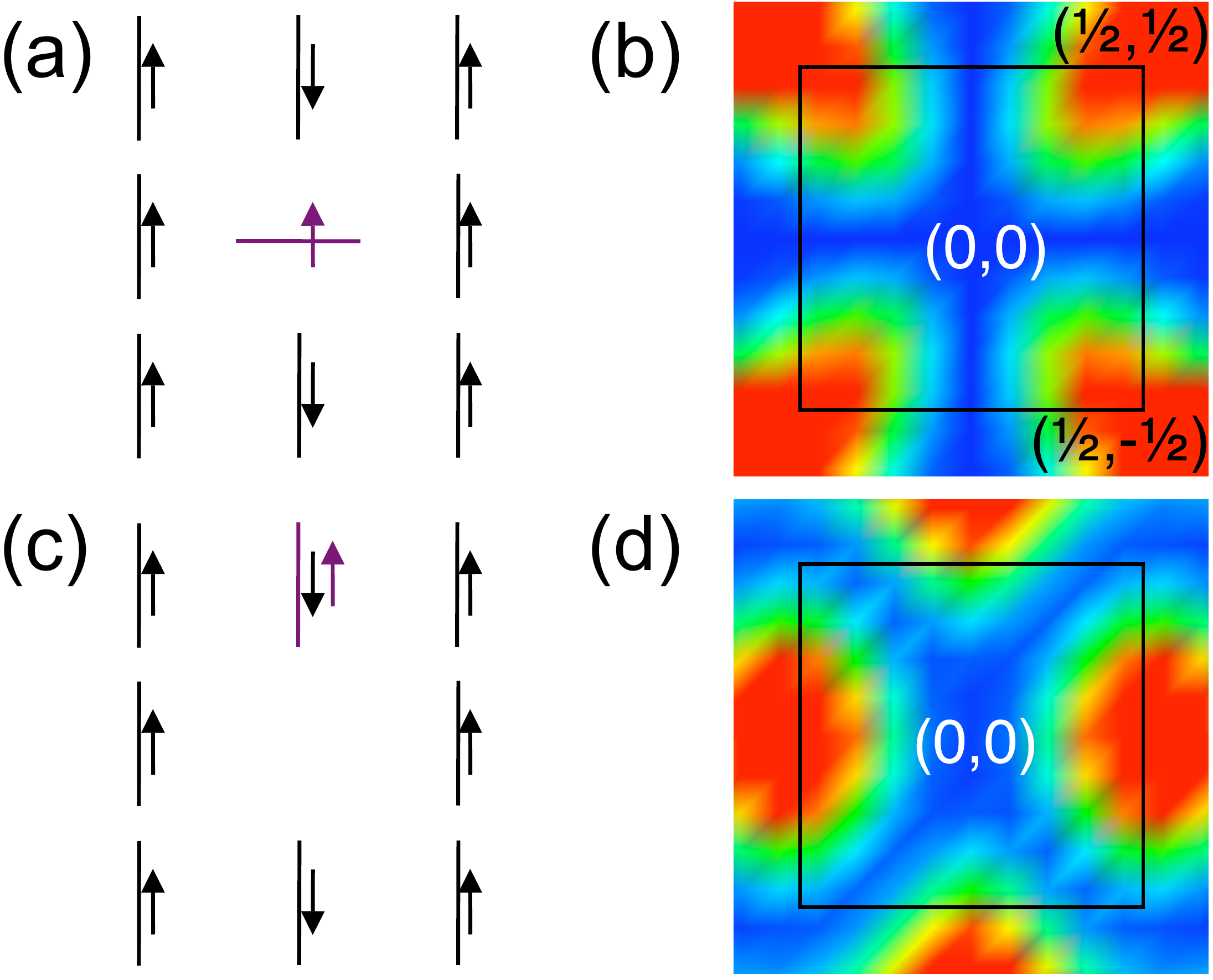}}
\caption{(color online).   (a) Cartoon of spin$+$orbital flip excitation discussed in text, assuming finite range spin and orbital correlations.  Vertical (horizontal) line denotes $d_{yz}$ ($d_{xz}$) orbital.  (b) Form factor for the excitation in (a).  Color map: low--medium--high = blue--green--red (dark--light--gray). (c)  Cartoon of inter-site spin-flip excitation.  (d)  Form factor map for the excitation in (c).} \label{fg:ff}
\end{figure}

Spin correlations characterized by ${\bf Q}\approx(\frac12,0)$, as observed here, correspond to the columnar pattern of Fig.~\ref{fg:latt}(b).
As shown in Ref.~\onlinecite{lee09}, such correlations are directly tied to the spin polarization of the $d_{yz}$ orbital and its ferro-orbital correlation.  [Similarly, spin correlations at ${\bf Q} \approx(0,\frac12)$ are tied to the $d_{xz}$ orbital.]   We propose that the spin-flip excitations in the present case are constrained by a strong coupling to orbital correlations, with 
two leading contributions illustrated in Fig.~\ref{fg:ff}.  
First, if we consider flipping a single spin, we find that the orbital also needs to flip from $d_{yz}$ to $d_{xz}$ [{\it c.f.} Fig.~\ref{fg:ff}(a)], in order to better utilize the superexchange energy.  (If the orbital remains unchanged, a local spin flip would have closed almost all the superexchange paths within the $d_{yz}$ and $d_{xz}$ subspace.)  This ``inter-orbital spin/orbital flip'' process introduces a strongly anisotropic form factor that goes to zero along the longitudinal direction, as shown in Fig.~\ref{fg:ff}(b).  This is consistent with the strong anisotropy of our neutron data and the previous observation at high energies.\cite{lums10}  
Another relevant process, one associated with itinerancy, is 
the ``inter-site spin flip'' process shown in Fig.~\ref{fg:ff}(c); the latter has a large form factor around $(\frac12, 0)$ [{\it c.f.} Fig.~\ref{fg:ff}(d)].  Interestingly, due to the columnar AF arrangement of the short-range spin correlations together with the Pauli exclusion principle, our proposed inter-site spin-flip process can only propagate along the $y$-direction at low-energy, in good agreement with the observed anomalous dispersion.  
A quantitative analysis of the dispersion (beyond the scope of this paper), will require evaluating the excitation energy as a function of wave vector, taking into account both processes.
In any case, the essence of our proposal is that the spin excitations are hybrids of the magnons and orbitons that have been postulated previously.\cite{krug09,lee09}

In conclusion, we have presented experimental evidence that the electronic and magnetic excitations of superconducting \fst\ are different from those of common band-structure and spin-wave models.  We have proposed that the intensity anisotropy of the magnetic excitations may be explained by considering coupled spin and orbital excitations.  If correct, this would imply that nematic excitations are a key feature of the normal state from which the superconductivity develops.  While the details of the magnetism are quite different,  the nematic correlations would provide an intriguing connection to the physics of cuprate superconductors.\cite{frad09}

We are grateful to S. A. Kivelson for helpful comments and to A. V. Fedorov for experimental assistance.  Work at the University of Virginia was supported by the Office of Science, US Department of Energy (DOE) through DE-FG02-07ER46384.  Work at Brookhaven is supported by the US DOE under Contract No.\ DE-AC02-98CH10886. PDJ and JMT are supported in part by the Center for Emergent Superconductivity, an Energy Frontier Research Center funded by the US DOE, Office of Basic Energy Sciences.  SPINS at NCNR is supported by the National Science Foundation under Agreement No. DMR-0454672.  ALS is operated by the US DOE under Contract No.\ DE-AC03-76SF00098


\end{document}